\documentclass{kapproc} 

\usepackage{procps} 

\usepackage[dvips]{graphicx}

\upperandlowercase

\kluwerbib 

\def\spose#1{\hbox to 0pt{#1\hss}}
\def\lta{\mathrel{\spose{\lower 3pt\hbox{$\mathchar"218$}}
     \raise 2.0pt\hbox{$\mathchar"13C$}}}
\def\gta{\mathrel{\spose{\lower 3pt\hbox{$\mathchar"218$}}
     \raise 2.0pt\hbox{$\mathchar"13E$}}}
\def\etal{{\it et al.\ }}
\begin{document}

\articletitle[On the Origin of S0 Galaxies]
{On the Origin of S0 Galaxies}

\author{Uta Fritze -- v. Alvensleben}
 
\affil{Universitätssternwarte Göttingen, Geismarlandstr. 11, 37083 Göttingen,
Germany}

\begin{abstract}
I will review the basic properties of S0 galaxies in the local Universe 
in relation to both elliptical and spiral galaxies, their neighbours on the 
Hubble sequence, and also in relation to dwarf spheroidal (dSph) galaxies. 
This will include colours, luminosities, spectral features, information about 
the age and metallicity composition of their stellar populations and globular 
clusters, about their ISM content, as well as kinematic signatures and their 
implications for central black hole masses and past interaction events, and the 
number ratios of S0s to other galaxy types in relation to environmental galaxy density. \\
I will point out some caveats as to their morphological discrimination 
against other classes of galaxies, discuss the role of dust and the wavelength 
dependence of bulge/disk light ratios. These effects are of importance for 
investigations into the redshift evolution of S0 galaxies -- both as individual 
objects and as a population. The various formation and transformation scenarios 
for S0 and dSph galaxies will be presented and confronted with the available 
observations.
\end{abstract}


\section[Introduction]{Introduction}
Until some years ago, S0 galaxies were thought to be simple and well-understood:
old passively evolving purely stellar systems in which everything had happened
in the distant past, which are boring now and bound to go on fading into a dark
and unspectacular future. In recent years a series of surprising observations
led to a vividly renewed interest in S0 galaxies, their formation,
transformation and evolution. 

\section[Conventional Wisdom and Recent Puzzles]{Conventional Wisdom and 
Recent Puzzles}
Located on the Hubble sequence between {\bf E}llipticals ({\bf E}s)
and {\bf Sp}irals ({\bf Sp}s) because of their strong r$^{1/4}-$bulges 
and their small and smooth
exponential disks with {\bf B}ulge-to-{\bf T}otal light ratios 
{\bf B/T}$\sim 0.6$ in the B-band, S0 galaxies hold keys to 3 fundamental
issues: {\bf 1.} does Hubble's classification 
describe a continuous sequence from Es through Sd and Irr galaxies with S0s
being the transition types or are Es and Sps fundamentally different 
and, if so, where do the S0s belong? {\bf 2.} What determines the properties of S0
galaxies: initial conditions or environmental effects, nature or nurture? {\bf
3.} The rate of evolution of S0 (and E) galaxies can test cosmological 
structure formation scenarios and constrain their parameters. 

Classically, S0 galaxies were thought to have formed in a monolithic initial
collapse with most of their {\large\bf S}tar {\large\bf F}ormation ({\bf SF}) 
on a short timescale in the early Universe. Their relatively unimportant stellar
disks (as compared to spirals) formed from leftover or accreted gas shortly
after the bulge as both components  
do contain red stellar populations, little gas and essentially no SF. S0
galaxies follow relations established for ellipticals, like the color--
magnitude relation, the black-hole mass -- velocity dispersion, 
Faber--Jackson, {\large\bf F}undamental {\large\bf P}lane ({\bf FP}), and
luminosity--metallicity relations. They also follow the Tully--Fisher relation
for spirals, although with larger scatter. The observations that S0s are
the dominant galaxy population in the central regions of nearby galaxy 
clusters while scarce in
clusters at intermediate redshift and in the field provided a major trigger into
renewed interest in the S0s and their formation or transformation histories. 
Recent detections of unexpected amounts of gas, dust and SF, kinematic
peculiarities, fine-structure, inward central color gradients, etc. added to
this new interest. Various scenarios have been suggested for the transformation
accompanying the infall into the gradually forming 
galaxy clusters 
of the spiral rich field galaxy population into the 
S0-, dSph-, and dE-rich population of todays rich clusters. Timescales for the
various aspects of the transformation have been discussed and possible
progenitors or early stages of these transformation processes are being looked
for in high and intermediate redshift clusters. Probably more than one
evolutionary path has led to the observed manifold of S0 and
dSph systems. 

\section[S0s on the Hubble Sequence]{S0s on the Hubble Sequence}
S0 follow many trends along the Hubble sequence originally defined by
morphological appearance, i.e. the bulge component getting less prominent and
the disk component and spiral structure getting more pronounced from early to
late-type galaxies: the colors of S0s are almost as red on
average as those of ellipticals and redder than those of spirals that get
increasingly bluer towards later Hubble types, their spectra, as well, have
properties that fit in between those of Es and Sa's. 
The content in neutral and
molecular gas in most S0s is low to negligible similar to ellipticals while it increases
systematically towards later spiral types. IRAS though has detected 50 \% of all E and S0 galaxies (Jura \etal 1987), later investigations revealed that cold gas is particularly common among peculiar S0s that also show relatively high HI content and H$_{\alpha}$. HI structures and kinematics often suggest an external origin. Dust masses from FIR measurements are typically a factor $\sim 10$ higher than those from optical extinction, indicating that the dust distribution is more extended than that of the star light. CO and HI are more strongly concentrated in the centers than for spirals (Bregman \etal 1998, Pogge \& Eskridge 1993, Sadler \etal 2000, Welch \& Sage 2003). However, even the most gas-rich S0s contain much less gas than expected if they had formed all their stars in an early burst and evolved as a closed-box since then (Faber \& Gallagher 1976). 
As opposed to massive ellipticals, 
S0s do not show luminous X-ray halos. Present-day star formation rates SFR$_o$ 
are low in
most S0s, their ratio of time averaged past SFR ${\rm \langle SFR \rangle}$ to 
present-day SFR$_o$ falls between the
respective values for Es and the earliest type spirals (e.g. Sandage 1986). 
Their SF efficiences, i.e. SFRs normalised to gas content, are higher than for spirals, for which they decrease further from Sa through Sd. In
terms of average luminosity and M/L$-$ratio, S0s also range between Es and Sa's.

All these observations indicate that the Hubble sequence is a continuous
sequence, not only in terms of B/T$-$light ratios, but also in terms of colors and
spectral properties, gas content, SFR$_o$, ${\rm \langle SFR \rangle}$, and, 
hence, stellar population ages. On the other hand, one might look at the S0s in
the light of the apparent dichotomy among ellipticals: Low-luminosity Es
are fast rotators, have power-law surface brightness profiles towards the
resolution limit, disky isophotes, and approximately isotropic velocity
dispersions, all compatible with the idea that dissipation has played a role in
their formation and dynamical evolution. Luminous Es, however, show slow or no 
rotation, are triaxial, have cores, boxy isophotes, and anisotropic velocity
dispersions (Kormendy \& Illingworth 1982). Kormendy \& Bender (1996) redefine the
Hubble sequence as extending from boxy Es via disky Es to S0s, Sa, ..., Sd, Irr. van den
Bergh (1994) argued in favor of the S0s sharing the dichotomy among the
ellipticals with faint S0s being flattened
and prolate and bright S0s being the ones that are truely intermediate between Es
and Sa's. Faber \etal's (1997) observations that, with one exception (NGC 524), 
S0s all have power-law surface
brightness profiles support this view. 
 
It has long been known that in terms of central parameters (central surface
brightness, central velocity dispersion) and absolute luminosity S0s and bulges
follow the relations among ellipticals while {\bf d}warf {\bf E}llipticals ({\bf
dE}s) and Globular Clusters follow different relations in
the famous Kormendy (1985) diagrams. From these diagrams and the knowledge
from stellardynamical modelling that mergers cannot increase the central phase
space density by more than a factor of two, it is clear that S0s or normal/giant Es
cannot be produced in mergers of dEs or dSphs. Ryden \etal (1999) found no distinct
structural differences between dEs and
dSphs/dS0s. 

S0s harbour nuclear black holes like bulges and Es, as revealed by absorption
or, in some cases, emission line analyses of high spatial resolution HST 
spectra. The two
S0s in Pinkney \etal's (2003) sample follow the black hole mass -- velocity 
dispersion relation M$_{\rm BH}-\sigma_{\rm e}$ for Es and bulges found by
Gebhardt \etal (2000) and Tremaine \etal (2002). If all S0s have central BHs
still seems to be an open question, e.g. in view of NGC 524's core. 

S0s follow the central Mg$_2~-~\sigma_o$ relation of ellipticals and have 
metallicity gradients shallower by factors 2$-$3 than spirals, 
therewith fitting into the trend of increasing metallicity gradients from early
to late-type spirals. In contrast to spirals and ellipticals, however, many S0s
apparently get older from inside out, i.e. their bulge stellar populations 
are younger than those of their disks. S0s follow the anticorrelation between
EW(H$_{\beta}$) and velocity dispersion $\sigma_o$ of ellipticals -- 
indicating that more massive galaxies have older stellar populations (Fisher \etal 
1996). The age of a
stellar population in this context means the time elapsed since the last epoch
of significant SF. Concerning the relative stellar population ages in the 
disks and bulges of field S0s, observations have revealed discrepant results:
BJHK photometry with surface brightness profile decomposition of 35 S0s showed
that because of similar $J-K$ and $H-K$ colors both components probably have
similar metallicities and the different $B-H$ colors indicate that disks are
$\sim 3-5$ Gyr younger than bulges. Some S0s show evidence for AGB light
contributions implying active SF until $2-3$ Gyr ago (Caldwell 1983, Bothun 
\& Gregg 1990). NGC 7332 which has emission line gas kinematically
decoupled from the stellar component and an A-type bulge spectrum (Bertola \etal
1992, Fisher \etal 1994, Hibbard \& Rich 1990), indicative 
of recent accretion and starburst events, is found by Bender \& Paquet (1999) 
to have a bulge star population spectroscopically older than its disk
population. NGC 5102 and NGC 404, resolved into individual 
stars with HST, both show blue inward color gradients and significant 
populations
of young stars $\gta 15$ and $\gta 300$ Myr, respectively, in their centers 
(Deharveng \etal
1997, Tikhonov \etal 2003). Investigations into stellar population ages of
33 S0s and 19 Es in the Coma {\bf cluster} over a wide range in luminosities $-20.5 \leq
{\rm M_B} \leq -17.5$ showed that 1. $\geq 40$ \% of the S0s (and none of the Es!)
had significant SF in their {\bf central} regions during the last 5 Gyr and that
2. the fraction of S0s with recent SF increases with decreasing luminosity
(Poggianti \etal 2001a). This luminosity dependence of the luminosity weighted
ages is independently found from optical and NIR photometry of S0s in Abell 2218
(z$=0.17$) by Smail \etal (2001). It explains the
discrepant results of earlier studies and support the idea of a dichotomy
between low and high-luminosity S0s.

S0s as well as ellipticals in Virgo, Coma and other nearby clusters -- both rich
and poor -- follow a
{\bf C}olor$-${\bf M}agnitude {\bf R}elation ({\bf CMR}) in the sense that 
luminous galaxies are
redder than fainter ones (Bower \etal 1992, Andreon 2003). This CMR is
primarily due to higher metallicities in brighter galaxies, as evidenced by
spectroscopic observations yielding similar relations between Mg$_2$ and
luminosity or Mg$_2$ and $\sigma$. In addition to metallicity differences, 
age differences may also contribute. E.g. Kuntschner \& Davis (1998) and
Kuntschner (2000) find from line strength analyses of Fornax Es
and S0s that the 
luminous Es \& S0s have formed their stars earlier or stopped their SF earlier
than the lower luminosity ones. The tightness of the CMR is
conventionally assumed to imply uniform old stellar population ages and, within
hierarchical galaxy formation scenarios, that more massive galaxies formed 
earlier from more massive building blocks. 

Schweizer \etal (1990) found fine structure, i.e. ripples, shells, plumes,
boxiness, X-structure, etc. in $\sim 50$\% of the {\bf field S0s}. 
The amount of fine structure as quantified by his fine structure parameter 
$\Sigma$ correlates with deviations from the CMR towards bluer
colors, with increasing H$_{\beta}-$ EWs, and with decreasing Mg$_2$ (Schweizer \&
Seitzer 1992) in the sense that $\sim 50$ \% of the field S0s had major 
mergers with significant starbursts about $3-8$ Gyr ago (Schweizer 1993, 1999). 

E/S0s form a homologous 3$-$parameter family with low scatter described by the
apparantly universal {\bf F}undamental {\bf P}lane ({\bf FP}) relation 
between effective radius ${\rm r_e}$, velocity dispersion
$\sigma$, and effective surface brightness ${\rm \langle I_e \rangle}$ within 
${\rm r_e}$ (Djorgovski \& Davis 1987,
Dressler \etal 1987, Jorgensen \etal 1993, 1996, Scodeggio \etal 1998a, b) 
\begin{center}   ${\rm log r_e = a \cdot log \sigma + b \cdot log \langle I_e
\rangle + c}$ \end{center}
The scatter is generally $\lta $ the measurement errors, smaller in clusters than
in the field and strongest for the low-luminosity S0s (van Dokkum \etal
2001). In combination with the Virial Theorem the FP relations imply a weak mass
dependence of the mass-to-light ratio ${\rm M/L \sim M^{0.2}}$ (Faber \etal 
1987). A salient feature of the FP is that it indicates an intimite relation
between spectral and dynamical evolution aspects of galaxies by coupling 
tightly structural and stellar population parameters. Observations of the FP in
the NIR show that 1.) the scatter is independent of passband, implying that
variations in age are balanced by variations in metallicity, and 2.) that the
slope increases steadily from U through K, pointing to systematic changes in
metallicity {\bf and} age {\bf and} DM content/homology breaking (Pahre \etal
1998).
Conventionally, the tight FP is interpreted in terms of E/S0s being very
homogeneous old stellar systems. However, it has been shown that
stellardynamical mergers conserve the FP and its small scatter (Capelato \etal
1995, Evstigneeva \etal 2004) and that E+Sp merger remnants soon after
merging come to lie on a {\it FP of changes} that is only tilted by $15^{\circ}$ 
relative
to the observed FP (Levine 1995). Observations by Lake
\& Dressler (1986) of a dozend well-known Sp -- Sp 
merger remnants from Arp's (1966) and Arp \& Madore's (1985) catalogues 
still featuring long tidal tails like NGC 7252 or NGC 3921 have
shown that -- very surprisingly -- these objects $\leq 1$ Gyr after their merger 
and strong starburst well fit into the L$- \sigma$- and 
FP$-$relations, casting serious doubt upon the idea that the very
existence and the tightness of these relations necessarily imply uniform old
ages for E and S0 galaxies. Apparently violent relaxation very fast transforms
the {\bf inner} regions within 1${\rm r_e}$.  

S0 galaxies not only follow relations for elliptical galaxies, they also obey 
the {\bf T}ully$-${\bf F}isher {\bf R}elation ({\bf TFR}) for spirals: 
${\rm L \sim V_{max}^{\alpha}}$ with $\alpha
\sim 4$ and ${\rm V_{max}}$ the maximum rotation velocity (Tully \& Fisher 1977, 
Pierce \& Tully 1992). Mathieu \etal (2002) find 6 S0s to follow the H-band TFR
with small scatter $\sim 0.3$ mag while being offset by $\sim 1.1$ mag to
fainter magnitudes in I, consistent with some fading after SF truncation (see
below). Results are still somewhat controversial with Hinz \etal (2001, 2003)
reporting a small offset $\leq 0.2$ mag from the H-band TFR and a large scatter
for massive (${\rm V_c \geq 200 km s^{-1}}$) S0s in Virgo and Coma and Neistein \etal (1999) 
finding a smaller offset in I and a large scatter for 18 nearby field S0s, in 
particular at the faint end. The
luminosity range of the sample, like for the stellar population ages, or the
field versus cluster environment might be a key to reconcile the apparently
discrepant results. Note that rotation curves are much more difficult to 
measure for S0s than for spirals 
because they largely have to rely on absorption lines. 
Moreover, about 50 \% of the Virgo 
S0s show weird rotation curves that do not allow to derive a V$_{\rm max}$ at
all (Rubin \etal 1999). Different formation paths may well lead to S0s with 
diverse properties. 

\section[Classification Problems]{Classification Problems}
S0s are distinguished against Sa galaxies either by their less prominent and
featureless disks or, in quantitative analyses, by some limiting B/T$-$light
ratio. Face-on S0s are hard to distinguish against ellipticals, the only
reliable method being long-slit absorption line spectroscopy that allows to
detect an S0's rapidly rotating disk. In a
magnitude-limited Coma sample Jorgensen \& Franx (1994) found only 12 \% Es
among the huge E/S0 population.  
For distant clusters often only the combined E/S0 population is considered. 
Using evolutionary synthesis models for bulge and disk components
separately with two different SF histories -- const. SFR as appropriate for disks and
exponentially declining SFRs (${\rm t_{\ast}\sim 1}$ Gyr) for bulges -- and 
combining
bulges and disks to obtain after a Hubble time the observed average B-band 
B/T$-$light
ratios of various Hubble types we could study both the wavelength and the
redshift dependences of B/T$-$light ratios (Schulz \etal 2003). We investigate
three different scenarios for the onset of bulge and disk SF: 1.) both start in the
early universe at z$\sim 3-5$, 2.) bulge SF starts at z$\sim 3-5$ and disk SF later
at z$\sim 1$, and 3.) disk SF starts at z$\sim 3-5$ and bulge SF later at z$\sim
1$.

\begin{figure}[ht]
\begin{center}
\includegraphics[width=0.49\linewidth]{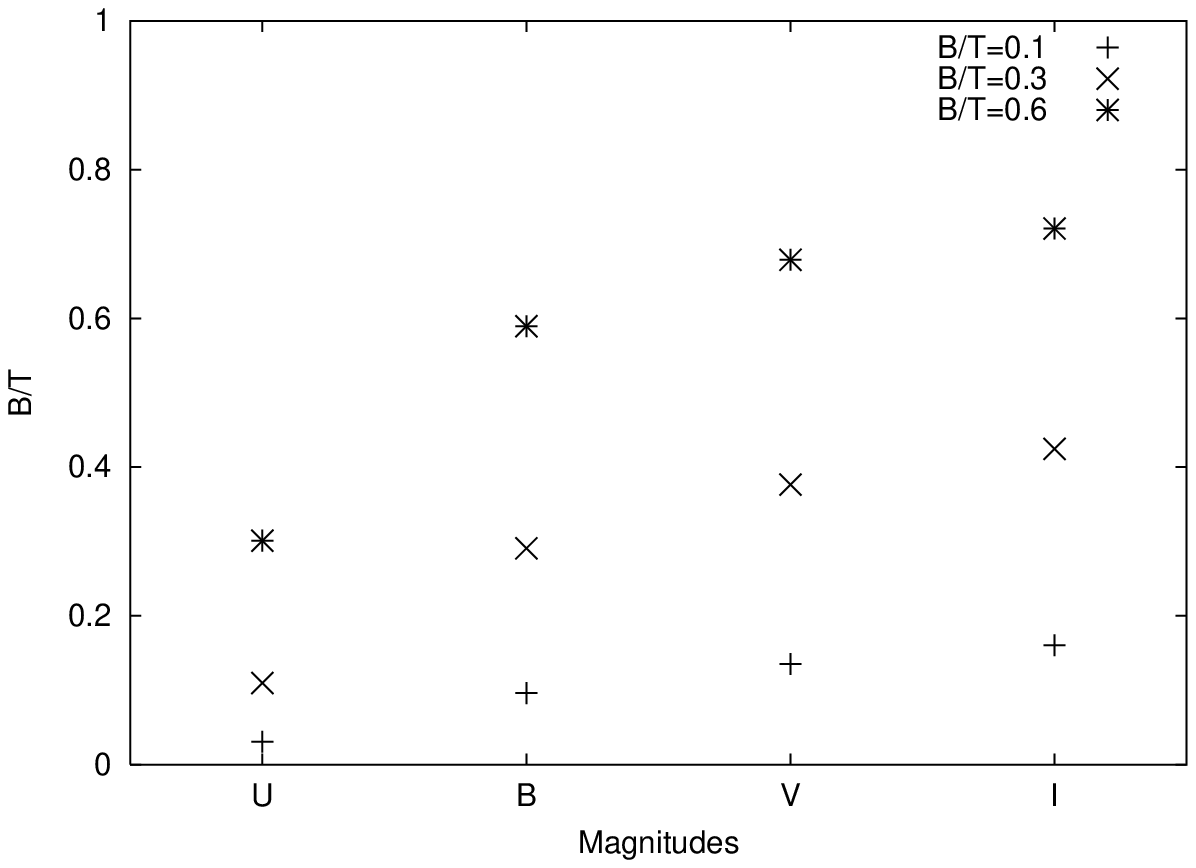}
\includegraphics[width=0.49\linewidth]{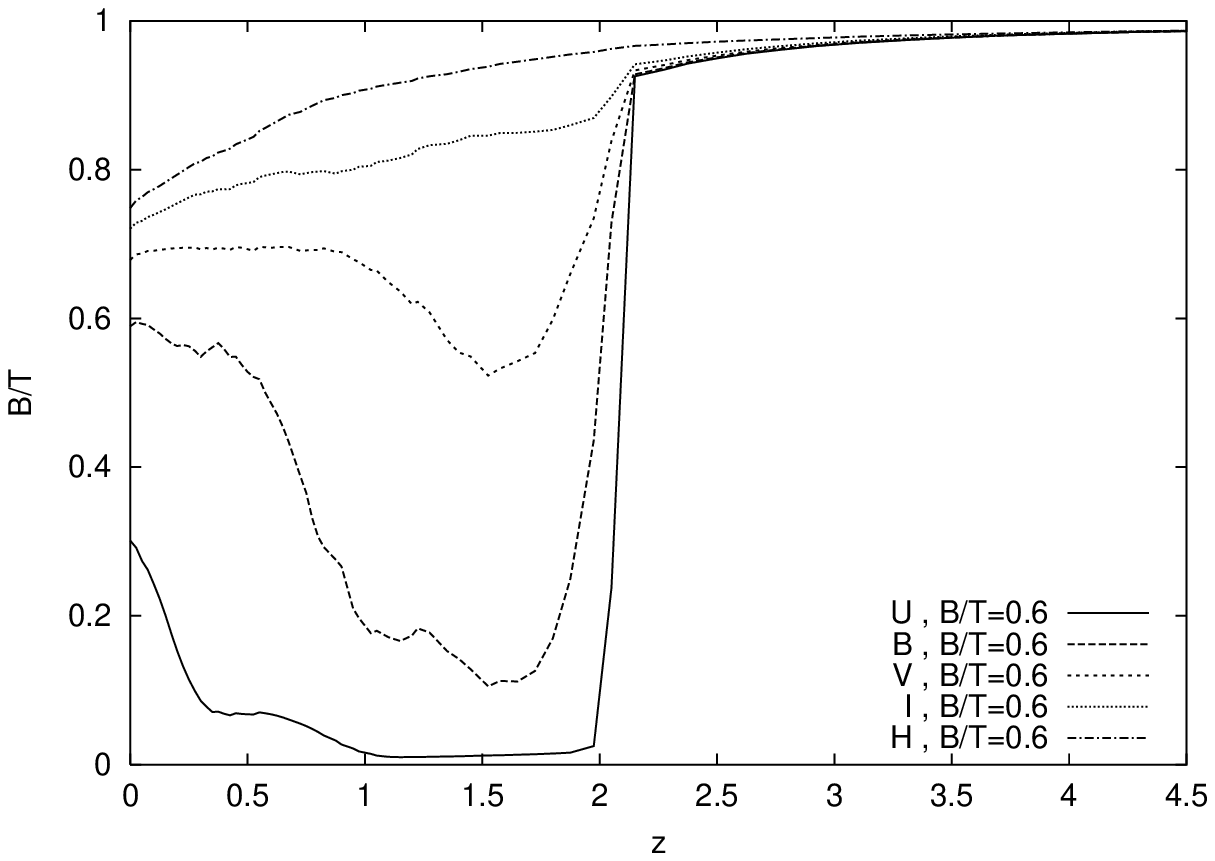}
\caption{(a) Wavelength dependence, (b) redshift
dependence of the B/T-light ratio for S0s. Bulge and disk SF both starting at ${\rm z \sim 3-5}$.}
\end{center}
\end{figure}

We found a significant wavelength
dependence of B/T$-$light ratios (Fig.1a): for S0s B/T increases by a factor $2-3$ from
$U$ through $K$ (for Sd's by a factor $3-4$) in agreement with H-band B/T$-$
determinations by Eskridge \etal (2002). The amount of increase slightly
depends on the relative ages of bulge and disk stars (cf. Schulz \etal for details). In addition, we found a
strong redshift evolution of B/T$-$light ratios that cannot be compensated for by
switching from $B$ to $V,~R,~I$ in a comparison with galaxies at redshifts
z$=$0.3, 0.5, 0.7. Models show that in any of the three scenarios B/T$-$light
ratios of S0s at higher redshift determined from surface profile decomposition 
in bands that correspond to restframe B are significantly overestimated. S0s at
higher redshift therefore have a high chance to be misclassified as Es (Fig.1b). 
The reason is the difference in time evolution of the disk
and bulge components due to their different SF histories. We therefore decided 
to give cosmological and evolutionary
corrections separately for bulge and disk components and the 3 scenarios that we
explored: bulge and disk of equal age, bulge older or younger than the disk (cf.
Schulz \etal 2003). It is true that changes get smaller towards longer
wavelengths and that therefore galaxy classification in I, as e.g. done for the HDF
galaxies by Marleau \& Simard (1998), or in a NIR band is less affected. For 
S0s and Es, unfortunately, the NIR B/T$-$light ratios are very similar and do
not allow to separate them from each other. 

\section[Formation and Transformation of S0s]{Formation and Transformation 
of S0s}
In the central regions of local rich virialised galaxy clusters S0 galaxies 
are the dominant galaxy type, making up as much as $\sim 60$\%
of the bright galaxy population. Dwarf galaxies (dEs, dS0s, dSphs) are about
twice as numerous as the luminous ones in Coma (de Lucia \etal 2004). Distant
clusters, on the other hand, contain increasing fractions of blue galaxies, as
first discovered by Butcher \& Oemler (1978, 1984). Van Dokkum (2001) 
and Dahlen et al. (2004) find a factor $\sim 5$ increase in the blue galaxy
fraction from z$\sim 0.5$ to z$=0$, i.e. over the last 5 Gyr (using ${\rm
H_o=65,~\Omega_m=0.3,~\Omega_{\lambda}=0.7}$). Most of the blue
galaxies in distant clusters are Sps and Irrs with ongoing SF (Smail \etal 
1997), some are having starbursts, others show Balmer absorption lines
indicative of recent starburst (Dressler \& Gunn 1983). Some of the red galaxies
also have strong Balmer absorption lines (E+A, k+a), i.e. are slightly older
post-starbursts (spectroscopic BO$-$effect). 
Fasano \etal (2000, 2001) and Couch
\etal (1998) show that while the fraction of ellipticals in the cluster 
population remains approximately constant from redshift ${\rm z \sim 0.5}$ to
${\rm z=0}$, the
fraction of spirals decreases by a factor $\sim 5$, and the fraction of S0s
increases by the same factor. This suggests a significant transformation of 
spirals into S0s from ${\rm z \sim 0.5}$ to
${\rm z=0}$, i.e. over the last 5 Gyr. At the
same time, the faint-to-luminous galaxy ratio in clusters increased 
from $\sim 1$ at ${\rm z=0.75}$ to $\sim 2$ at ${\rm z=0}$ (de Propris \etal 2003). 

The field galaxy population also being spiral-rich and poor in S0s and Es, the
transformations of spirals into S0s and of luminous into dwarf galaxies seem 
to be linked with processes during the
continuous infall of field spirals towards increasingly rich clusters. 
Formation and
evolution of galaxy clusters and Large Scale Structure hence appear to go
hand in hand with morphological and spectral transformation of important 
parts of their galaxy population.

A large variety of formation and transformation scenarios have been proposed for
S0s over the years and it seems that more than one of them is needed to account
for the observed diversity of the S0 population. Which one(s) is (are) realised or
prevalent may depend on environment and epoch; massive and low-mass S0s may have
different formation and evolution histories.

Like bulges in later-type spirals the spheroidal componets of S0s could have 
formed in the classical {\it fast initial collapse with rapid SF} scenario 
-- at least for those 50\% of the field S0s that
do not show any fine structure or signs of central rejuvenation. 

Internal instabilities in spiral disks can form bars that efficiently funnel
 gas and stars into the central parts and bars, in turn, can dissolve to form 
 a bulge component. The huge bulges of S0s require several
circles of bar formation and destruction as seen in numerical simulations
(Combes \etal 1990, Raha \etal 1991). Depending on the relative amounts of gas
and disk stars funneled into the central regions, stellar populations in the
bulge might be older or younger than those in the disk. 

Major spiral -- spiral mergers have been shown to result in elliptical galaxies
since the Toomres' pioneering work. Incomplete violent relaxations transforms 
stellar disks into de Vaucouleurs profiles while saving small gradients. The
delayed and protracted backfall of HI from the tidal tails onto the main
body, as e.g. observed in 
NGC 7252, can rebuild a gaseous disk on timescales of few Gyr that by and by 
transforms into a stellar disk, resulting in a disky elliptical or
an S0 galaxy as seen in combined N-body and hydrodynamical simulations by
Hibbard \& Mihos (1995) and in Barnes' (2002) models. Few Gyr after the global or nuclear starbursts 
triggered respectively in prograde and retrograde encounters (Bekki 1995) the
merger remnants have been shown to reach typical S0 -- and somewhat later
elliptical -- galaxy colors and spectra in our evolutionary synthesis models
(Fritze -- v. Alvensleben \& Gerhard 1994a, b). It is clear that this major
merger scenario is only viable for massive and luminous S0s. Only in field or
group environments galaxy encounter velocities are low enough for efficient 
merging. Hence the luminous S0s in clusters must have been preprocessed in the
field or within infalling groups -- in agreement with observations that 
{\bf luminous} 
S0s have stellar population ages as old as ellipticals. In this scenario, the stellar population in
the bulge will be older than that in the disk. 
Observations of the rich cluster MS 1054-03 at z$=0.83$ reveal a substantial 
number of major mergers, 17\% in total, not only in the center but also in
the outer parts, consistent with the idea of enhanced merging in infalling groups (van Dokkum \etal
1999). Gavazzi \etal (2003) observationally caught a collective starburst among 
the members of a galaxy group falling into the cluster Abell 1387. 

Minor mergers (e.g. mergers with galaxy mass ratios 3:1) or accretion events  
have been shown by Barnes (1996) and Bekki (1998) to be viable routes towards intermediate or
low-luminosity S0s, again in the field or in groups rather than in dense
clusters where galaxies have too high relative velocities. The nature and gas
content of the objects involved determines the outcome. A Sp+dE or a 
Sp+(d)Irr merger will result in an S0 with a bulge older or younger than 
the disk depending on the strength of a possibly triggered starburst. 

Major mergers among gas-rich galaxies form populous systems of new star clusters, many of which are compact and strongly enough bound to survive for Gyrs (Schweizer 2002, Fritze -- v. Alvensleben 1998, 1999). Their enhanced metallicities and younger ages, when determined from spectroscopy or multi-wavelength photometry, are much more precise tracers of past violent SF events than the integrated galaxy light (Anders \etal 2004a, Fritze -- v. Alvensleben 2004a, b). If the young star clusters forming in minor mergers are also long-lived still is an open question (Anders \etal 2004b, Fritze -- v. Alvensleben 2004b). S0 galaxies with bimodal {\bf G}lobular {\bf C}luster ({\bf GC}) color distributions in any case testify back to a major merger origin. Both the luminous S0 NGC 1380 (Kissler -- Patig \etal 1997) and the low-luminosity S0 NGC 3115 show bimodal GC color distributions. The field stars in NGC 3115 also show the same bimodal color distribution (Elson 1997). Kundu \& Whitmore (2001) estimate that $\gta 10-20$ \% of the S0s have bimodal GC color distributions and, hence, a clear major merger origin. Age differences between the blue metal-poor and the red metal-rich GCs are small ($\lta 3$ Gyr). Hence, the mergers producing the red GC population in the starbursts they triggered must have happened early. 
On average, S0s have less populous GC systems than ellipticals of comparable luminosity and are closer to spirals in this respect, although with large scatter.  

Harassment is an important process in dense cluster environments, where fast
galaxy -- galaxy encounters destabilise the disks of infalling spirals, 
drag out strong short-lived tidal tails and tear away stars from their 
outer disks, leaving the inner parts as dE, low-luminosity S0 or dSph 
galaxies on timescales
of few Gyr while releasing up to 50\% of the original stellar mass of the
incoming spiral to the cluster potential (Moore \etal 1995, 1998). As
the remnants of this process consist of the former central parts of much more
massive galaxies, they are to be expected to deviate from the Mg$_2-\sigma~-$
relation towards too high metallicities for their mass. Poggianti \etal
(2001b) report a significant population of low-luminosity galaxies in Coma with
exactly these properties: enhanced metallicities for their luminosities. 
Apart from being harassment products they could
also be Tidal Dwarf Galaxies, i.e. recycling galaxies forming out of gas and
stars torn out into a tidal tail from a disk galaxy in interaction and
becoming self-gravitating and dynamically independent there (cf. Duc \& Mirabel
1999, Weilbacher \etal 2002, 2003).

Beyond these different types of galaxy -- galaxy interactions the presence of a
hot dense ICM as observed at X-wavelengths in the centers of nearby rich 
clusters will affect the gaseous components of infalling spirals. Processes like
ram pressure stripping or sweeping of their HI disks have been predicted as 
the density and pressure of the hot ICM are observed to be comparable or higher 
than those of the ISM within galaxies, and they can directly be seen in terms 
of increasingly truncated and displaced HI disks in spirals towards the center
of the Coma cluster (Cayatte \etal 1990, Bravo -- Alfaro \etal 2000). It is
clear that SF in these anemic spirals (van den Bergh 1976) will also be
truncated. If the instability intrigued by the rapid removal of considerable
amounts of HI will lead to a starburst consuming all the molecular gas and part of the central HI in
one shot before final SF truncation and if it can also affect the stellar
configuration is less clear. Observations that the galaxy population in clusters
starts to deviate from the field galaxy population in terms of their SF
activity at unexpectedly large
distances from cluster centers (around ${\rm 3-4~R_{vir}}$) (Lewis \etal 2002,
Gomez \etal 2003, Balogh \etal 2004, Gerken \etal 2004) raised the idea 
that the ICM densities that far out might still be high enough to drive away the
low density gas from infalling galaxy halos that otherwise could serve as a 
reservoir for disk accretion and, hence, for SF (Larson \etal 1980, Bekki \etal 2002). If realised by nature, this halo gas
stripping or starvation would lead to SF strangulation on a rather long timescale of order Gyr. 

We have investigated the effect of SF truncation without or after a preceeding
starburst on the photometric and spectral evolution of various spiral types 
in Bicker \etal (2002). Using evolutionary synthesis models that successfully
describe undisturbed galaxies of types Sa, Sb, Sc, and Sd, we added bursts of
various strengths and/or SF truncation at various evolutionary ages. 
Figs. 2a, b show the time evolution of $(B-V)$ colors and of B-band
luminosities M$_{\rm B}$
for starbursts and/or SF truncation occuring at ${\rm z=1}$ in a 6 Gyr 
old Sc galaxy. Unless 
the SFR does not go to zero, all models reach S0 galaxy colors
fairly rapidly: within 2 Gyr after SF truncation and within $4-5$ Gyr after
a starburst, depending on its strength, in case of $B-V$. The luminosity evolution, of
course, strongly depends on whether or not a starburst occured before SF
truncation: between pure SF truncation and strong burst models there is 
a difference in final luminosity of 2 mag in M$_{\rm B}$. 

\vskip-.3cm
\begin{figure}[ht]
\begin{center}
\includegraphics[width=0.49\linewidth]{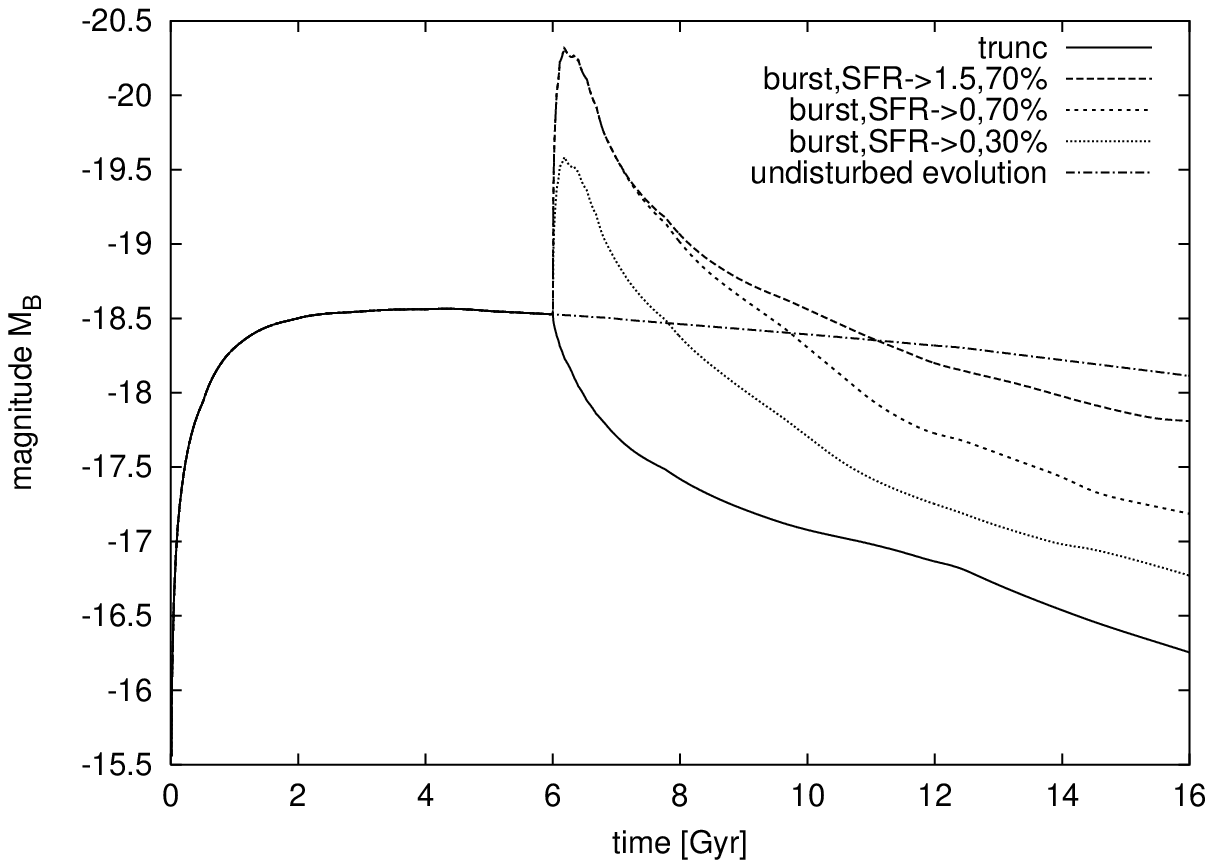}
\includegraphics[width=0.49\linewidth]{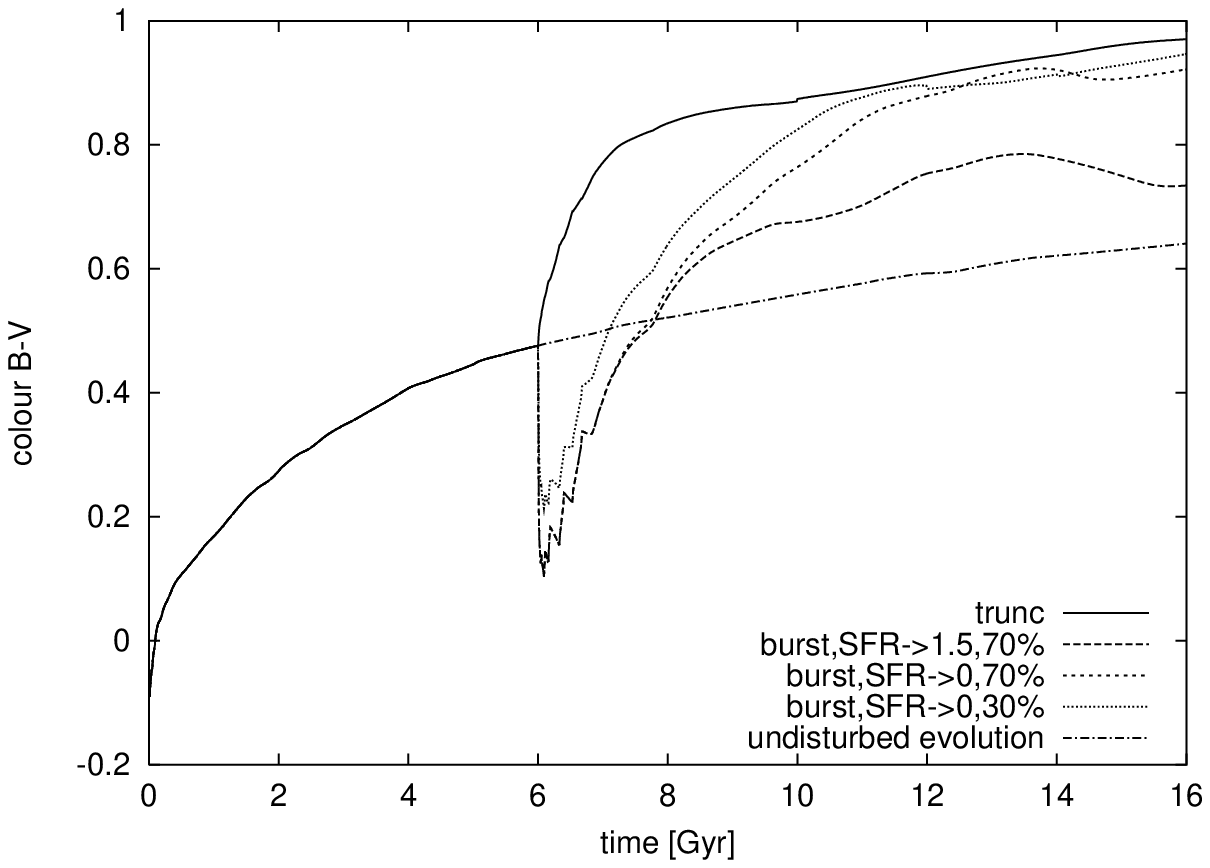}
\includegraphics[width=0.49\linewidth]{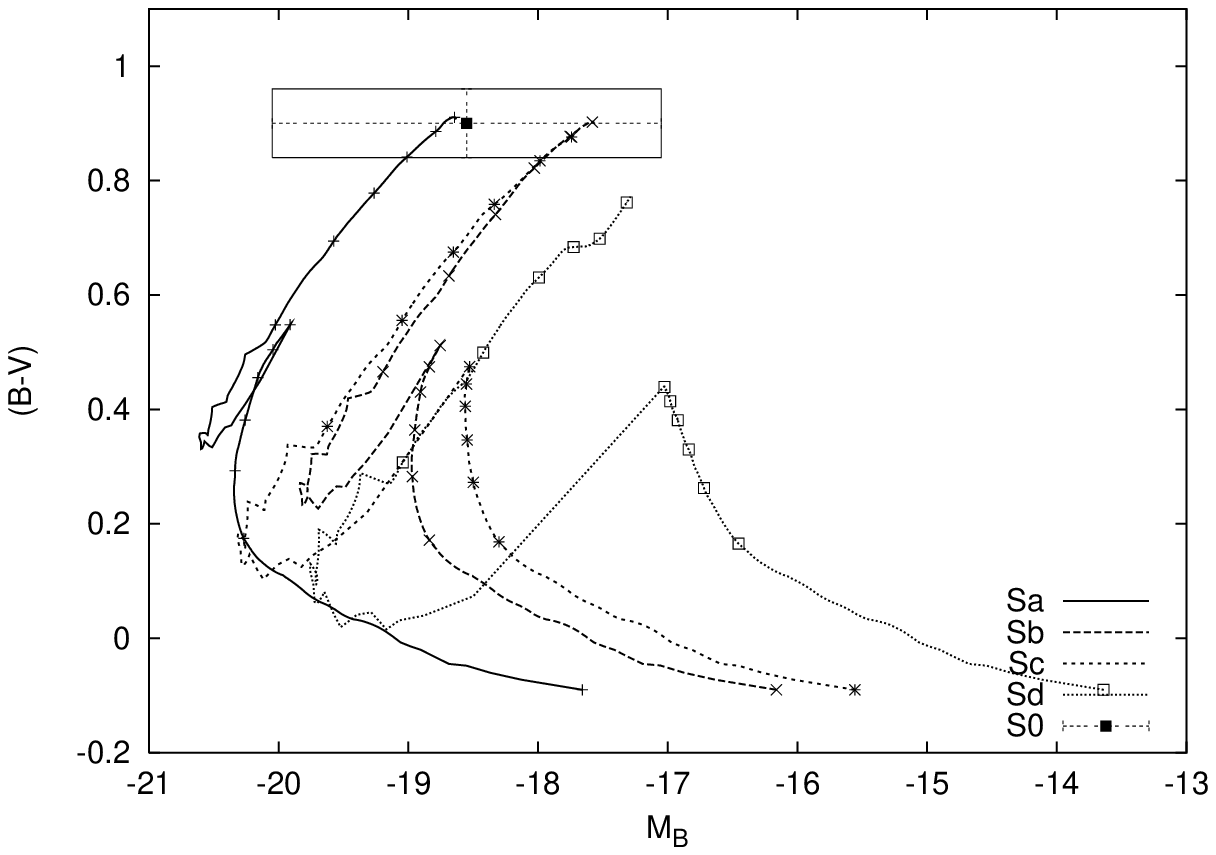}
\includegraphics[width=0.49\linewidth]{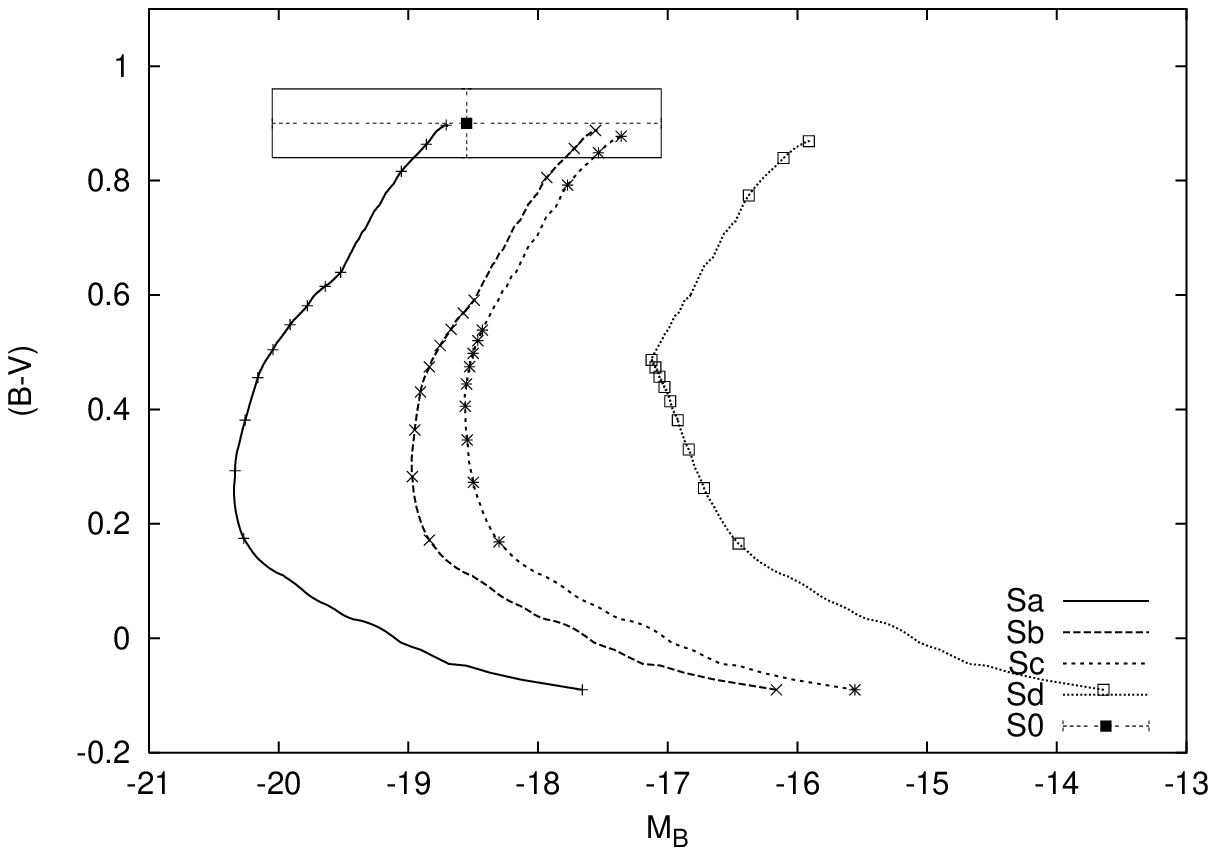}
\caption{Luminosity (a) and color (b) evolution of an Sc spiral model with starbursts and/or SF truncation occuring at an age of 6 Gyr. CMD for various spiral models with starbursts occuring at 6 Gyr (c) and with SF truncation at 9 Gyr (d). Ticks mark 1 Gyr steps.  The box gives average observed S0 color and luminosity with 1$\sigma$ ranges.}
\end{center}
\end{figure}
\vskip-.3cm

The
photometric evolution in $UBVRIJHK$ of these models is compared with observed
S0 properties in terms of {\bf C}olor $-$ {\bf M}agnitude {\bf D}iagrams ({\bf
CMD}s). Note that our undisturbed models are calibrated in mass as to yield 
after a Hubble time the average observed M$_{\rm B}$ for the respective spiral
type. As an example, Figs. 2c, d show in a CMD the evolution of various spiral types with
strong bursts and/or SF truncation occuring at ages of 6 Gyr and 9 Gyr,
corresponding to redshifts ${\rm z=1}$ and ${\rm z=0.5}$, respectively. 
The box in
the upper left indicates the observed average color and luminosity of S0s with
their 1$\sigma$ ranges. 
About 4 Gyr after their strong starbursts Sa, Sb, Sc models all reach the 
color range 
of S0s, Sd models remain too blue. Only the Sa model reaches the average
luminosity of S0s. Sb and Sc models after strong starbursts are able to
account for the fainter S0s. Only 1.5 Gyr after SF truncation without a
preceeding starburst all models 
show S0 galaxy colors, but again only the Sa progenitors reach average 
S0 luminosities,
the Sb snd Sc progenitors turn into low-luminosity S0s, and Sd galaxies with
truncated SF end up with luminosities of dSphs. SF truncation in 3 Gyr young 
Sa and Sb progenitors, i.e. at ${\rm z=2}$, would produce galaxies too red 
for S0s. At such young
ages, however, the dense hot ICM of today's rich cluster centers that is 
thought
to be responsible for SF truncation by sweeping out the gas from infalling
spirals might not yet have accumulated. We also considered
mergers of equal spiral types that double the mass in stars and gas and the
luminosity while not changing the colors. After strong bursts accompanying these
mergers, Sa progenitors reach the bright end of the S0 galaxy luminosity
distribution, and even Sd-galaxies, when merging with each other, reach average 
S0 luminosities. As far as the color evolution is concerned, our results are 
confirmed by an independent study by Shioya \etal (2002). Including also the
luminosity evolution, however, our models allow for tighter constraints on the
manifold of possible S0 progenitors than
theirs. To investigate the recently renewed suggestion that SF strangulation by halo gas
starvation might truncate SF on a longer timescale of $\geq 1$ Gyr (Larson
\etal 1980, Bekki \etal 2002), we
recalculated models under these conditions and we also analyse the spectral
evolution, e.g. the evolution of emission and absorption lines (Fritze \& 
Bicker {\sl in prep.}). 

We find that
SF truncation on a short timescale of order $10^8$ yr is followed by a phase
of moderate H$_\delta -$ strength, while SF strangulation on a timescale of
$10^9$ yrs does not develop enhanced H$_\delta$ absorption. SF truncation
after a preceeding burst results in an H$_\delta -$ strong phase of $\sim 1.5$
Gyr duration (see also Barger \etal 1996, Poggianti \etal 1999, and Shioya
\etal 2004). Note that red 
H$_\delta -$ strong galaxies are also called E+A or k+a types.  

We conclude that the progenitors of low-luminosity S0s can be Sa through Sc galaxies
that experienced a starburst $>3$ Gyr ago, as well as Sa/Sb galaxies with SF
truncation at ages between 6 and 9 Gyr, i.e. at ${\rm 1 \leq z \leq 0.5}$, 
or Sc galaxies with SF truncation as late as ${\rm z=0.5}$, i.e. after 9 Gyr 
of undisturbed evolution. The strength of the burst does not
make much difference. The progenitors of luminous S0s can only be early-type 
spiral -- spiral or multiple mergers with starbursts occuring at ages $\leq 9$ Gyr, i.e. at ${\rm z \geq 0.5}$. 

The progenitors of low- and high-luminosity
S0s that had experienced a starburst have gone through an H$_\delta -$ strong 
phase, first blue and then red, of $\sim 1.5$ Gyr duration. The progenitors of those low-luminosity
S0s and dSphs that had their SF truncated without a starburst can, at maximum,
have gone through a phase of intermediate-strength Balmer absorption provided
the SF truncation happened on a short timescale of order $10^8$ yr.   
 
Barger \etal (1996) and Kodama \& Smail (2001) quantitatively investigated the
possibility of a spiral-to-S0 transformation in terms of galaxy numbers and
morphologies. Couch \& Sharples	(1987) report a  
high proportion ($\sim 30$\%) of actively SFing and starbursting galaxies in 
3 rich clusters at z$=0.31$, many of which show signs of interactions 
while still having disk morphologies, and a high fraction of H$_{\delta}-$ 
strong galaxies, most of which are regular spheroids, indicating that the 
timescale for the photometric transformation is somewhat shorter than that 
for the morphological transformation, but that the strong Balmer absorption 
features after strong starbursts still are detectable after the 
morphological transformation is accomplished. Maybe in mergers which are probably the origin of strong starbursts the morphological transformation is faster than for other transformation processes? 
	
The decrease of the blue galaxy fraction towards ${\rm z=0}$ probably is a 
consequence of the interplay between a
decline in the galaxy infall rate with cosmic time, the overall 
decrease in gas content
and SFR in the field galaxy population, and the increasing cluster richness
together with the 
successive build-up of their
dense hot ICM content.  
While at intermediate and high redshift, the starburst and E+A galaxies in
clusters were bright high-mass objects, the actively SFing, starburst and
postburst 
galaxies in local clusters are predominantly low-luminosity late-type dwarf
galaxies. This trend is called {\it downsizing} effect (Bower \etal 1999,
Poggianti 2004) and also reported for the
field galaxy population (Cowie \etal 1996). Duc \etal (2002) conclude from ISOCAM mid-IR observations that up to 90\% of the SF in a cluster ${\rm z=0.18}$ is hidden. If this would selectively apply for the most massive and metal-rich galaxies it might explain an apparent downsizing effect simply by a strong prevalence of dust at low redshift.

\section[Redshift Evolution of S0 Galaxy Populations]{Redshift Evolution of 
S0 Galaxy Populations}
Van Dokkum \etal (1998) investigate the {\bf redshift evolution of the CMR} for 194
E/S0 galaxies in DL 1358+62 at z$=0.3$. They find the scatter, i.e. the age
spread, for Es very small
and independent of clustercentric radius ${\rm R_{cl}}$. The scatter for the 
S0s is equally small in the center, however, it increases considerably towards 
larger ${\rm R_{cl}}$ with the S0s being increasingly offset towards bluer
colors at large ${\rm R_{cl}}$. This shows that while ellipticals have
terminated their SF well before accretion, S0s stop SF in the outer parts of the
cluster. Stanford \etal (1998) extend this study to 19 clusters at z$\sim 0.9$
using NIR photometry and find the local slope conserved, the scatter, i.e. the
apparent age spread among E/S0s still small and the degree of evolution
independent of cluster richness or ${\rm L_X}$. The fact that the slope for
these young galaxy populations is not different from the local one implies that
it results from a correlation of galaxy mass with metallicity, rather than
with age. 

Studies of the {\bf redshift evolution of the FP} aim at 3 aspects: the redshift
evolution of its slope tells about size and mass evolution, i.e. about
accretion and evolution in orbital structure, changes in the zero-point
indicate evolution in M/L, due in part to the fading of the stellar population
and possibly also affected by rejuvenation through accretion and SF, the
redshift evolution of the scatter is a measure of the homogeneity of the E/S0
population at various lookback times. Semianalytical $\Lambda$CDM structure
formation models predict that high density environments should collapse first
and fastest. Hence, they lead us to expect cluster galaxies to be older than field
galaxies and therefore to expect differences in the redshift evolution of 
the FPs of field
and cluster galaxies (Kauffmann 1996, Diaferio \etal 2001). Remember again that age in this context means time since 
the last epoch of significant SF.  

The redshift evolution of the FP for {\bf cluster E/S0s} in the redshift interval 
${\rm 0.3 \leq z \leq 0.8}$ as studied by Kelson \etal (1997), Bender \etal
(1998), van Dokkum \etal (1998), Jorgensen \etal (1999), and Wuyts \etal (2004)
indicates that the scatter remains close to local until ${\rm z \sim 0.5}$ and only
increases at ${\rm z=0.8}$, the zero-point shift indicates $\sim 1$ mag 
brightening to ${\rm z=0.8}$. The evolution in M/L is slow, implying a 
galaxy formation redshift ${\rm z_f>2.8}$ consistent with results from 
the redshift evolution of the CMR. 
The redshift evolution of the FP for {\bf field galaxies} was studied
by means of Keck-LRes spectroscopy 
by Gebhardt \etal (2003) on 21 Es and 15 S0s from the DEEP sample in the
redshift range ${\rm 0.3 \leq z \leq 1}$ with ${\rm \langle z \rangle =0.8}$ and
by van Dokkum \& Ellis (2003) on 10 E/S0s from the HDF-N with ${\rm 0.56 \leq z
\leq 1.02}$. Both groups find slope and scatter similar to local, and a 
zero-point offset by 2.4 mag in $I$ (corresponding to rest-frame $B$ at ${\rm
z=1}$), marginally in agreement with passive evolution of simple old populations. 
Comparison with the evolution of cluster E/S0s indicates an age difference
between cluster and field E/S0s of less than 2 Gyr, which poses a problem for
hierarchical structure formation models that predicted a larger age difference. 
 
The {\bf redshift evolution of the TF relation} remains inconclusive at
present for the S0s due to the difficulties in measuring stellar velocity
dispersions from absorption lines in high redshift spectra.  

\section[Conclusions and Open Questions]{Conclusions and Open Questions}
Located between ellipticals and spirals on the Hubble sequence, S0 galaxies play
a key role in 3 fundamental and still largely open questions: as to the nature
of Hubble's sequence, to the respective roles of nature versus nurture in
shaping galaxy properties, and as valuable test cases for cosmological galaxy
formation scenarios. Are S0s transition types on one continuous line from 
E through Sd galaxy types or do they belong to either of two fundamentally 
different classes 
of objects? I reviewed arguments for and against boths options and conclude that
current evidence seems to favor a dichotomy among S0s similar to the one for
ellipticals. In this sense, the luminous S0s have formed their stars earlier
than the low-luminosity S0s and may have had different formation scenarios.  

S0 galaxies share many properties with other dynamically hot stellar systems, as
e.g. the relation between central black hole mass and velocity dispersion, the
color--magnitude and Fundamental Plane relations, while also following the
Tully--Fisher relation for spirals. We have seen that constraints from the FP
and CMR on the ages and formation histories of S0s are less tight than
originally thought, and that ~50\% of the field S0s show evidence for
important intermediate age stellar contributions in their central regions
together with morphological and kinematic traces of interactions events.
Detections of HI, CO, HII, dust and SF activity in quite a number of S0s adds
to this view. 
Together with the strong redshift evolution of the S0 and spiral fractions in
galaxy clusters this provides evidence that nurture rather than nature, environmental
effects rather than initial conditions have shaped S0 galaxies, that they are
{\bf transformation} rather than formation products. It is not clear at present
if any classical {\it early collapse + short timescale SF} S0s do at all exist in
the field -- for sure not in clusters. It also seems clear that more than one
transformation scenario must be at work to produce the observed manifold of S0s.
Major spiral--spiral/Irr mergers with starbursts may result in luminous S0s, as
shown by dynamical simulations as well as evolutionary spectral synthesis, with
the protracted backfall of HI from the tidal tail(s) leading to the build-up of
a secondary stellar disk on timescales of ~3 Gyr. Minor mergers (3:1) or
accretion events on the one hand, harassment, tidal stripping of stars and
ram-pressure sweeping of gas on the other hand may result in lower-luminosity
S0s and dSphs. Any kind of merging or accretion requires low relative velocities
and, hence, is more probable within groups
before or during their infall into clusters or in very early stages of cluster
evolution. It may leave kinematic
peculiarities, fine structure, and positive color gradients that can survive for a
few Gyrs. If the bursts are strong enough the resulting new star cluster
population with its age and metallicity may be better suited to trace back 
the SF history of its parent galaxy than the integrated
light. Pixel-by-pixel analyses provide spatially resolved information about
the respective contributions of old and younger stars. 
Harassment through fast galaxy--galaxy encounters is probably the
dominant process in dense clusters, its end products will be low-luminosity
S0s, dSphs or dEs that can be expected to deviate from the
luminosity--metallicity relation because they are the leftover central parts of
originally much more massive galaxies. Interactions with the dense hot X-ray
emitting ICM in nearby rich galaxy clusters is observed to efficiently remove HI
disks from spirals plunging into it, cutting their gas supply and, hence,
truncating their SF while destabilising the stellar disks. The timescale for 
the morphological transformation seems to
be longer than that for the photometric transformation, the strong Balmer absorption features after strong bursts, however, seem to survive the morphological transformation, at least in some cases. 

If the much less dense halo gas is removed from spirals by the same process
at larger cluster-centric radii already -- with the effect of cutting the SFing disk
off from its accretion reservoir causing SF strangulation on a relatively long
timescale seems less clear. Detailed statistics of the various progenitor and
transition types (normal and strong emission line galaxies, blue and red
H$_{\delta}-$ strong galaxies) in
clusters of varying richness and degree of relaxation and at various
cluster-centric radii should ultimately allow to identify the relative importance
of all these different formation paths. The fact that the morphology--density
relation seems to be continuous from the densest cluster centers out to the
field probably indicates that, beyond the direct interaction with the clusters'
central ICM and potential, more localised processes, e.g. in groups, must also 
play an important role. 

It
seems for sure that the S0 galaxies and the questions as to their origin will
keep us excited for years to come and may still hold further
surprises.  

\begin{acknowledgments}I gratefully acknowledge generous travel support from the
organisers of this conference without which I could not have attended.
\end{acknowledgments}

\end{document}